\documentclass[aps,prd,twocolumn,groupedaddress,preprintnumbers,nofootinbib]{revtex4}
\usepackage{url}

\usepackage[latin1]{inputenc}
\usepackage[english]{babel}

\usepackage{amssymb,amsmath,amsthm,cancel,hyperref,graphicx,xcolor}
\usepackage{picinpar,graphicx,xypic}
\usepackage{booktabs}
\usepackage{mathrsfs}
\usepackage[font=small,labelfont=bf,margin=3mm,labelsep=period,tableposition=top]{caption}
\usepackage[a4paper, top=2.2cm, bottom=2.2cm, left=2cm, right=2cm, bindingoffset=0mm]{geometry}

\numberwithin{equation}{section}
\pdfoutput=1

\newcommand{\pr}{$^\prime$}

\newcommand{\mh}{m_H}
\newcommand{\mt}{m_t}

\newcommand{\sss}{\scriptscriptstyle\rm}

\newcommand{\Lum}{\mathscr{L}}

\newcommand{\as}{\alpha_s}
\newcommand{\muf}{\mu_{\sss F}}
\newcommand{\muh}{\mu_{\sss H}}
\newcommand{\mus}{\mu_{\sss S}}
\newcommand{\mur}{\mu_{\sss R}}

\let\originalleft\left
\let\originalright\right
\renewcommand{\left}{\mathopen{}\mathclose\bgroup\originalleft}
\renewcommand{\right}{\aftergroup\egroup\originalright}

\def\beq{\begin{equation}}
\def\eeq{\end{equation}}
\def\({\left(}
\def\){\right)}
\def\[{\left[}
\def\]{\right]}
\allowdisplaybreaks

\graphicspath{{images/}}

\begin{document}

\title{The three loop soft function for N$^3$LL$^\prime$ gluon fusion Higgs production in SCET}

\author{Marco Bonvini}
\author{Luca Rottoli}
\affiliation{Rudolf Peierls Center for Theoretical Physics, 1 Keble Road\\
  University of Oxford, OX1 3NP Oxford, UK}

\preprint{OUTP-14-19P}


\begin{abstract}
We derive the three loop soft function for inclusive Higgs production in gluon fusion,
and use it to perform the resummation of the Higgs cross section at N$^3$LL$^\prime$ in SCET.
We improve the accuracy of the resummation by including contributions of collinear origin.
We include finite top, bottom and charm mass effect where available.
These results are available through the public code \texttt{ResHiggs}. 
\end{abstract}

\maketitle

\section{Introduction}

Although not directly measurable, the inclusive cross section for Higgs production
is an observable of great interest, both experimentally (serving as normalization for
exclusive distributions) and theoretically, due to the bad convergence properties of its
perturbative expansion in the strong coupling $\as$.
At Large Hadron Collider (LHC) energies and for the Standard Model (SM) Higgs,
the inclusive cross section is largely dominated by the gluon fusion production mode,
where the Higgs couples to the gluons through a fermion loop, prevalently a top quark.
Being the top quark heavier than the SM Higgs, an effective field theory can be introduced
where the top is integrated out, leading to an effective pointlike gluon-gluon-Higgs vertex.

QCD corrections to this process are known for a long time at next-to-leading order (NLO)~\cite{NLO}
and next-to-next-to-leading order (NNLO)~\cite{NNLO} in the heavy top limit; the exact result at NLO is also known~\cite{NLOmt},
while finite top mass corrections at NNLO have been computed as an expansion in $\mh/\mt$,
being $\mh$ the Higgs mass and $\mt$ the top mass~\cite{NNLOmt}.
The effective theory turns out to predict well $K$-factors (ratios to the LO cross section)
for current LHC energies, but the accuracy of the theory is expected to decrease when the collider
energy increases, due to the wrongly predicted high energy behavior at parton level~\cite{Marzani:2008az}.

These QCD corrections are huge: the NLO correction amounts to about $120\%$ of the LO cross section,
while NNLO correction adds another $\sim80\%$.
On top of that, canonical renormalization and factorization scale variation bands (by a factor of $2$ about the Higgs mass)
do not overlap, as a symptom of a very bad convergence.
Choosing a smaller central renormalization scale ($\mh/2$) slightly improves the described behavior,
with NLO and NNLO bands now overlapping, but still being far from a decent convergence.

For these reasons, there has been a big effort in going beyond NNLO.
Within the heavy top effective theory, the full next-to-next-to-next-to-leading order
(N$^3$LO) computation is ongoing~\cite{Gehrmann:2010ue,NNNLO,Anastasiou:2014vaa,NNNLO-NS};
in the meantime, some approximate predictions based on known all-order behaviors
appeared~\cite{Ball:2013bra,Bonvini:2014jma,deFlorian:2014vta}.
Alternatively, higher order contributions can be predicted
by all-order resummations, specifically the so called $\pi^2$ resummation~\cite{PI_SQ} and
soft gluon resummation~\cite{deFlorian:2012yg,Bonvini:2014joa,Ahrens:2008nc}.

Soft gluon resummation aims to resum logs of $1-z$,
being $z=\mh^2/\hat s$, with $\hat s$ the partonic center of mass energy;
these logs are large at partonic threshold, where $z\to1$.
Although the Higgs at LHC is very far from physical threshold, the partonic threshold
is still dominant, due to the shape of the gluon luminosity that favors larger values
of $z$ in the convolution with the partonic cross section~\cite{Ahrens:2008nc,Bonvini:2012an}.
However, whether the soft logarithms dominate at LHC strongly depends on the actual form
of such logarithms, differing from each other by subdominant contributions vanishing at $z=1$,
as discussed at length in Refs.~\cite{Ball:2013bra,Bonvini:2014joa,Bonvini:2010tp}.

Soft gluon resummation is currently known at
next-to-next-to-next-to-leading logarithmic accuracy N$^3$LL$^\prime$~\cite{Bonvini:2014joa},
where the prime denotes the inclusion of some formally higher order terms,
which however contain the bulk of the next order correction~\cite{Bonvini:2014qga}.
This result has been computed with the standard formalism of QCD, sometimes
referred to as direct QCD (dQCD). Alternatively, soft resummation can be realised
within soft-collinear effective theory (SCET), where it is known up to N$^3$LL~\cite{Ahrens:2008nc}.

The purpose of this paper is to extend the SCET result to N$^3$LL$^\prime$.
To do this, we need the hard and soft functions at three loop order, one order higher than what is used
in the N$^3$LL result of Ref.~\cite{Ahrens:2008nc}. While the first ingredient is known since a few years~\cite{Gehrmann:2010ue},
the second was unknown at the time of writing.\footnote{Shortly before the completion of this letter,
this missing ingredient has been computed in Ref.~\cite{Li:2014afw}.}
Exploiting the equivalence of SCET and dQCD~\cite{Bonvini:2014qga,Sterman:2013nya,Bonvini:2012az,Bonvini:2013td,Ahrens:2008nc},
we extract the three loop expression of the soft function from the N$^3$LL$^\prime$ dQCD result. 
We also consider a modification of the form of the soft logarithms that includes
important collinear contributions, and 
discuss the comparison to the dQCD result of Ref.~\cite{Bonvini:2014joa}.

\section{Soft gluon resummation in SCET}

\begin{table*}[t]
\begin{center}
\begin{tabular}{llcccl}
  Notation\pr & Notation* & $\Gamma_{\rm cusp}$, $\beta$ & $\gamma_i$ & $H$, $\tilde s_{\rm Higgs}$ & Natural matching\\
  \hline
  LL        &   LL      & 1-loop & --- & tree-level & LO\\
  NLL       &   NLL*    & 2-loop & 1-loop & tree-level & LO\\
  NLL\pr    &   NLL     & 2-loop & 1-loop & 1-loop & NLO\\
  NNLL      &   NNLL*   & 3-loop & 2-loop & 1-loop & NLO\\
  NNLL\pr   &   NNLL    & 3-loop & 2-loop & 2-loop & NNLO\\
  N$^3$LL   &   N$^3$LL*& 4-loop & 3-loop & 2-loop & NNLO\\
  N$^3$LL\pr&   N$^3$LL & 4-loop & 3-loop & 3-loop & N$^3$LO
\end{tabular}
\caption{Orders of the logarithmic approximation and of the ingredients needed to achieve it.}
\label{tab:count}
\end{center}
\end{table*}

The inclusive Higgs cross section at a hadron collider with center of mass energy $\sqrt{s}$
can be written as a sum over partons of convolutions
\beq
\sigma = \sigma_0\sum_{i,j}\int_{\mh^2/s}^1\frac{dz}{z} \Lum_{ij}\(\frac{\tau}{z},\muf^2\) C_{ij}(z,\mh^2,\muf^2)
\eeq
of a parton luminosity
\beq
\Lum_{ij}\(x,\muf^2\) = \int_{x}^1\frac{dx'}{x'} f_i\(\frac{x}{x'},\muf^2\) f_j\(x',\muf^2\),
\eeq
which is itself a convolution of parton distribution functions (PDFs) $f_i(x,\muf^2)$,
and a perturbative partonic coefficient function $ C_{ij}(z,\mh^2,\muf^2)$;
the prefactor $\sigma_0$ is chosen such that $C_{gg}$ is normalised to 1 at LO.

In SCET, hard, collinear and soft modes are integrated out at subsequent matching steps,
leading to a factorised form for the partonic coefficient function. In the inclusive case,
the coefficient function for the $gg$ channel can be written as~\cite{Ahrens:2008nc}
\beq\label{eq:scetfact}
C_{gg}(z,\mh^2,\muf^2) = H(\muf^2) S(z,\muf^2),
\eeq
where $H(\muf^2)$ is a hard function and does not depend on the Higgs kinematics
and $S(z,\muf^2)$ is a soft function, and we are omitting an implict $\mh$ dependence in each factor.
In Ref.~\cite{Ahrens:2008nc}, the hard function is further factorised due to the usage of the
heavy top effective theory; here, instead, we keep finite top mass dependence (up to NNLO),
as done for instance in Ref.~\cite{Berger:2010xi}.

The hard and soft functions obey evolution equations in the energy scale $\muf$, whose solution can be written
in a closed form~\cite{Ahrens:2008nc,Berger:2010xi}: this can be used to write each function at
a hard scale $\muh$ and a soft scale $\mus$ respectively,
simply supplementing Eq.~\eqref{eq:scetfact} by evolution
factors from each scale to the common scale $\muf$.
Using the results of Ref.~\cite{Ahrens:2008nc}, we write
\begin{align}\label{eq:scetfact2}
&C_{gg}(z,\mh^2,\muf^2) = H(\muh^2) U(\muh^2,\mus^2,\muf^2)\\
&\qquad\times\tilde s_{\rm Higgs}\(\log\frac{\mh^2}{\mus^2}+\partial_\eta,\mus^2\)
\frac{z^{-\eta}}{(1-z)^{1-2\eta}}  \frac{e^{2\gamma\eta}}{\Gamma(2\eta)},\nonumber
\end{align}
where $\eta = 2 a_{\Gamma_{\rm cusp}}(\mus^2, \muf^2)$ and
\begin{align}
&U(\muh^2, \mus^2, \muf^2) = \frac{\as^2(\mus^2)}{\as^2(\muf^2)}
\left[ \frac{\beta(\as(\mus^2))/\as^2(\mus^2)}{\beta(\as(\muh^2))/\as^2(\muh^2)} \right]^2 \nonumber \\
&\quad\times\left|\left(\frac{-\mh^2 - i \epsilon}{\muh^2} \right)^{-2 a_{\Gamma_{\rm cusp}}(\muh^2,\mus^2)} \right| \\
&\quad\times \left|\exp[4 S(\muh^2,\mus^2)-2 a_{\gamma^S} (\muh^2, \mus^2)+ 4 a_{\gamma^B}(\mus^2,\muf^2)]\right|,\nonumber
\end{align}
having introduced the definitions
\begin{align}
S(\nu^2, \mu^2) &= - \int_{\as(\nu^2)}^{\as(\mu^2)} d \alpha \frac{\Gamma_\textrm{cusp}(\alpha)}{\beta(\alpha)}
\int_{\as(\nu^2)}^\alpha \frac{d \alpha'}{\beta(\alpha')}, \nonumber\\
a_\gamma(\nu^2, \mu^2) &= - \int_{\as(\nu^2)}^{\as(\mu^2)} d \alpha \frac{\gamma(\alpha)}{\beta(\alpha)}.
\end{align}
The anomalous dimensions $\Gamma_{\rm cusp}$, $\gamma^S$, $\gamma^B$ and the QCD $\beta$ function
should be included according to Tab.~\ref{tab:count} to obtain a given logarithmic accuracy.
Up to N$^3$LL, all these ingredients can be found in Ref.~\cite{Ahrens:2008nc}, except the unknown 4-loop
$\Gamma_{\rm cusp}$, which is usually estimated with a Pad\'e approximant~\cite{Moch:2005ba}
(and has a negligible impact on the final result).
To reach N$^3$LL\pr\ we need in addition the three loop hard and soft functions;
the first has been computed in Ref.~\cite{Gehrmann:2010ue}, and the latter is presented in the next Section.

In Tab.~\ref{tab:count} we also show different ways of counting the logarithmic accuracy
(unprimed and primed, or, using a notation introduced in Ref.~\cite{Bonvini:2013td}, starred and unstarred),
depending on the order of the matching functions $H$ and $\tilde s_{\rm Higgs}$. A natural logarithmic counting
at the exponent would lead to the unprimed (starred) accuracies; however, it is known~\cite{Bonvini:2014qga}
that the inclusion of the next order matching functions, leading to the primed (unstarred) accuracy,
contains the bulk of the next logarithmic order correction. Also, it is natural to match the N$^k$LL result
to N$^{k-1}$LO, while the N$^k$LL\pr\ result is naturally matched to N$^{k}$LO, because the
hard and soft function contain ingredients to the same order.
However, since currently the N$^3$LO is not complete, in our analysis we will match N$^3$LL\pr\ to NNLO,
which effectively corresponds to estimating the fixed order $\as^3$ contribution with the
order $\as^3$ expansion of the N$^3$LL\pr\ resummed result.

\section{The three loop soft function}

To derive the three loop soft function, we use the fact that the
expansion of the N$^k$LL\pr\ must match the soft part of the fixed order N$^k$LO result.
This is a direct consequence of the way SCET is constructed: the functions $H$ and $\tilde s_{\rm Higgs}$
are found by subsequent matchings of full QCD onto SCET.
Alternatively, this can be understood in terms of the equivalence of SCET and
dQCD~\cite{Bonvini:2014qga,Sterman:2013nya,Bonvini:2012az,Bonvini:2013td,Ahrens:2008nc}:
for a particular choice of the scales $\mus$ and $\muh$, SCET and dQCD coincide to all orders in $\as$.
Additionally, the expansion of the SCET N$^k$LL\pr\ result does not depend on $\mus$ and $\muh$ up to order $\as^k$:
so, up to this order, the expansion must coincide with the same expansion in dQCD, independently on the scale choice.

Therefore, to determine the soft function $\tilde s_{\rm Higgs}$ to order $\as^3$, we expand the SCET result
Eq.~\eqref{eq:scetfact2} to order $\as^3$ and equate it to the soft part of the fixed N$^3$LO result~\cite{Anastasiou:2014vaa}
(or, equivalently, the third order expansion of the dQCD resummed N$^3$LL\pr\ result~\cite{Bonvini:2014joa}).
Since to this order the scale choice is immaterial, we are free to set $\muh=\mus=\muf$,
which switches off the resummation and makes the computation easier, since $U=1$
and $\eta=0$.
However, before setting $\eta=0$ we need to compute the derivatives in the first argument of $\tilde s_{\rm Higgs}$,
and regularise with a plus distribution the $z\to 1$ behavior.
In fact, it turns out to be easier to go to Mellin space first,
since the coefficient function is an ordinary function in $N$ space, and the $\eta\to0$ limit is harmless.
We have carried out the computation in \texttt{Mathematica}.
Writing the expansion of $\tilde s_{\rm Higgs}$ as
\begin{widetext}
\begin{align}
\tilde s_{\rm Higgs}(L,\as) &= 1+ \frac{\as}\pi \tilde s_{\rm Higgs}^{(1)}(L)
+ \(\frac{\as}\pi\)^2 \tilde s_{\rm Higgs}^{(2)}(L)
+ \(\frac{\as}\pi\)^3 \tilde s_{\rm Higgs}^{(3)}(L) +\ldots
\end{align}
we find for the third order coefficient
\begin{align}
\tilde s_{\rm Higgs}^{(3)}(L) &=
C_A^3 \bigg[\frac{L^6}{48}-\frac{11 L^5}{144}+
   \left(\frac{925}{1728}-\frac{\zeta_2}{16}\right) L^4
   + \left(\frac{11
   \zeta_2}{144}+\frac{7 \zeta_3}{8}-\frac{1051}{648}\right) L^3
   + \left(-\frac{13 \zeta_2^2}{80}-\frac{67 \zeta_2}{288}-\frac{209
   \zeta_3}{144}+\frac{20359}{5184}\right) L^2
\nonumber\\ & \qquad\qquad
   + \left(\frac{11
   \zeta_2^2}{40}+\zeta_2
   \left(\frac{193}{648}-\frac{\zeta_3}{24}\right)+\frac{1541
   \zeta_3}{216}-3 \zeta_5-\frac{297029}{46656}\right) L
\nonumber\\ & \qquad\qquad
   +\frac{11657
   \zeta_2^3}{15120}-\frac{4261 \zeta_2^2}{2160}+
   \frac{23333 \zeta_2}{46656}-\frac{11 \zeta_2\zeta_3}{24}+\frac{67
   \zeta_3^2}{36}-\frac{21763 \zeta_3}{3888}-\frac{121
   \zeta_5}{72}+\frac{5211949}{839808}\bigg]
\nonumber\\ &
+C_A^2 n_f
   \bigg[\frac{L^5}{72}-\frac{41 L^4}{432}+
   \left(\frac{457}{1296}-\frac{\zeta_2}{72}\right) L^3
   + \left(\frac{5
   \zeta_2}{144}+\frac{\zeta_3}{72}-\frac{793}{864}\right) L^2
   + \left(\frac{\zeta_2^2}{20}-\frac{19 \zeta_2}{648}-\frac{113
   \zeta_3}{216}+\frac{31313}{23328}\right) L
\nonumber\\ & \qquad\qquad
+\frac{389
   \zeta_2^2}{2160}-\frac{1633 \zeta_2}{23328}+\frac{19
   \zeta_3}{81}-\frac{\zeta_5}{12}-\frac{412765}{419904}\bigg]
\nonumber\\ &
+C_A C_F n_f \bigg[ \frac{L^3}{48}
  + \left(\frac{\zeta_3}{4}-\frac{55}{192}\right) L^2
  + \left(-\frac{\zeta_2^2}{10}-\frac{19\zeta_3}{36}+\frac{1711}{1728}\right) L
  + \frac{19\zeta_2^2}{180} + \frac{\zeta_2\zeta_3}{12}-\frac{55 \zeta_2}{576}
  +\frac{355\zeta_3}{648}+\frac{7\zeta_5}{18}-\frac{42727}{31104}\bigg]
\nonumber\\ &
+C_A n_f^2
   \bigg[\frac{L^4}{432}-\frac{5 L^3}{324}+\frac{25 L^2}{648}+
   \left(-\frac{\zeta_3}{18}-\frac{29}{729}\right)L+\frac{13
   \zeta_2^2}{1080}-\frac{\zeta_2}{648}+\frac{55
   \zeta_3}{972}-\frac{4}{6561}\bigg].
\label{eq:stilde3}
\end{align}
\end{widetext}
Eq.~\eqref{eq:stilde3} is the main analytic result of this letter.%
\footnote{Shortly before the completion of this letter, Ref.~\cite{Li:2014afw}
published a direct Feynman diagram computation of the same object:
the two results agree.}
It is of course the same in the exact theory and in the heavy top effective theory,
since this difference is all contained in the hard function $H$.

\section{\boldmath The Higgs cross section at N$^3$LL$^\prime$}

We are now ready to present the results for the N$^3$LL\pr\ inclusive Higgs cross section in gluon fusion.
To facilitate the comparison with Ref.~\cite{Bonvini:2014joa}, we use the NNLO set of parton distributions NNPDF 2.3~\cite{NNPDF23}
with $\as(m_Z)=0.118$ at each order, and $\as$ running at three loop order.
We set the top mass to $\mt=172.5$~GeV, and we take the bottom and charm masses from
the PDF set for consistency, namely $m_b=4.75$~GeV and $m_c=1.41$~GeV.
Regarding the fermion mass dependence, we use the same setup as in Ref.~\cite{Bonvini:2014joa},
namely the common LO prefactor is exact with top, bottom and charm running in the loop,
the NLO correction is also exact, and the NNLO contains finite top mass corrections~\cite{NNLOmt}.
Since the effect of the fermion masses is included in the hard function $H$ in SCET,
we do include the same effects in $H$ at the same orders, while for the third order we use the heavy top limit value.
We use the public code \texttt{ResHiggs}~\cite{ResHiggs}, where we have implemented
the resummation in the SCET formalism.

\begin{figure*}[t]
  \includegraphics[width=0.45\textwidth,page=1]{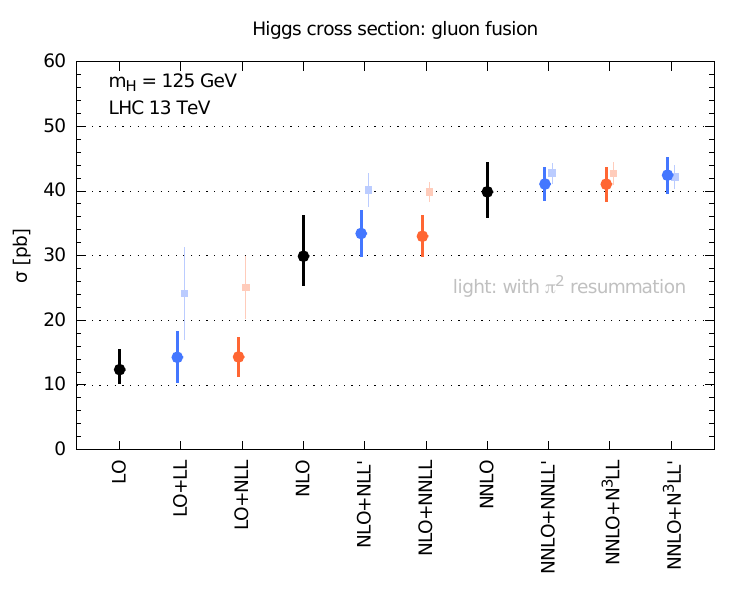}
  \qquad
  \includegraphics[width=0.45\textwidth,page=2]{SCET_hadr_xsec_res_125_13_uncertainty.pdf}
  \caption{Higgs cross section at fixed order and at resummed level for $\mh=125$~GeV and LHC with $\sqrt{s}=13$~TeV.
    In the left plot we use the same choice for the soft logarithms adopted in Ref.~\cite{Ahrens:2008nc},
    while in the right panel we include a collinear improvement, as described in the text.
    Resummed results including $\pi^2$ resummation are shown in lighter colors.
    Fixed order results are computed using the code \texttt{ggHiggs}~\cite{ResHiggs}.}
  \label{fig:1}
\end{figure*}

In Fig.~\ref{fig:1} (left) we show the cross section at different fixed and resummed orders
for $\mh=125$~GeV and LHC at $\sqrt{s}=13$~TeV.
The error band represents scale variation only, and is computed differently for the fixed order results
and for the resummed results.
Specifically, at fixed order we use canonical scale variation by a factor of $2$ up and down
with respect to the central scale, taken to be the Higgs mass $\mh$, for the factorization and renormalization
scales $\muf$ and $\mur$ independently, and requiring additionally $1/2<\mur/\muf<2$.
For the resummation, we follow the procedure of Ref.~\cite{Ahrens:2008nc}, namely we vary each
of the scales $\muf$, $\muh$ and $\mus$ about their central values, and add the resulting errors in quadrature
(after symmetrising each error).
The central values for $\muf$ and $\muh$ are the Higgs mass $\mh$, and the variation is by a factor of $2$ up and down.
For $\mus$, the central value is the average of two scales, called $\mus^{\rm I}$ and $\mus^{\rm II}$
in Ref.~\cite{Ahrens:2008nc}, obtained by requiring that the one loop contribution of $\tilde s_{\rm Higgs}$
to the cross section is $15\%$ of the total or at minimum, respectively; the variation is then performed
by letting $\mus$ vary between $\mus^{\rm I}$ and $\mus^{\rm II}$.

It is clear from the plot that the inclusion of resummation at each order has a small impact
on the predicted cross section, given that each resummed result lies within the error band of the
fixed order result to which it is matched. Even the new NNLO+N$^3$LL\pr\ prediction, the last point of the plot,
is compatible with the NNLO, although the central value is slightly larger, in agreement with the tendency found in
Refs.~\cite{Ball:2013bra,Bonvini:2014jma,deFlorian:2014vta,Bonvini:2014joa}.
Note also that the scale variation bands at resummed level are smaller than the fixed order bands at each order (except LO),
hinting that the uncertainty is underestimated, given that this is already the case for the fixed order results.

These phenomena are in fact easy to understand. The choice of the soft logarithms adopted in Ref.~\cite{Ahrens:2008nc}
and used in this plot corresponds to plus distributions of the form $(1-z)^{-1}\log^k\frac{1-z}{\sqrt z}$.
The presence of the $\sqrt z$ factor in the logarithm comes from kinematics, and, though subleading at large $z$,
improves the agreement of the resummation when expanded in powers of $\as$ with the fixed order;
however, the fixed order expansion of the resummation systematically underestimates the full result~\cite{Ball:2013bra},
leading to such small impact to all orders.
In Ref.~\cite{Ball:2013bra}, on top of this kinematic improvement,
it is shown that it is possible to include a class of subdominant
(i.e., suppressed by powers of $1-z$, and therefore beyond the leading power accuracy of SCET)
logarithms of collinear origin to all orders in $\as$~\cite{Kramer:1996iq,Contopanagos:1996nh,Catani:2001ic}.
The simplest realization of this collinear improvement consists in multiplying all the plus
distributions by a factor of $z$, which is equivalent in $N$ space to a shift $N\to N+1$~\cite{Ball:2013bra,Bonvini:2014joa}.
With the inclusion of collinear contributions, adopted also in Ref.~\cite{Bonvini:2014joa},
the agreement with fixed order improves significantly.

Therefore, we modify the soft logarithms with this additional $z$ factor, and show the results in
Fig.~\ref{fig:1} (right). Note that we consistently change the values of $\mus^{\rm I}$ and $\mus^{\rm II}$,
which we recompute according to the prescription of Ref.~\cite{Ahrens:2008nc} described above.
As expected, the contribution of the soft logarithms is larger with this choice, also leading to a larger
(and more realistic) scale uncertainty, mostly driven by $\muf$ uncertainty.
Also, the convergence is improved: excluding the LO+LL result (which contains to few information),
the NLO+NLL\pr, NNLO+NNLL\pr\ and NNLO+N$^3$LL\pr\ are compatible within the respective uncertainty bands.
Also, the convergence with primed accuracies seems to be better than the one of the unprimed results,
in agreement with Ref.~\cite{Bonvini:2014qga}.

The cross section at NNLO+N$^3$LL\pr\ including the collinear contributions is predicted to be $\sigma = 47.5\pm 6.7$~pb,
corresponding to an increase of $19\%$ with respect to the fixed NNLO.
This is compatible with the result of Ref.~\cite{Bonvini:2014joa}, which is $\sigma = 48.2^{+4.2}_{-2.0}$~pb
for the same configuration, thereby confirming the observation~\cite{Bonvini:2014qga} that,
practically, resummation performed in dQCD and SCET with a soft scale choice as in Ref.~\cite{Ahrens:2008nc}
leads to similar results when equivalent forms of the soft logarithms are adopted.

Finally, we have also checked that the inclusion of the $\pi^2$ resummation, adopted as default in Ref.~\cite{Ahrens:2008nc},
does not affect the final NNLO+N$^3$LL\pr\ result significantly (by less than $1\%$), giving $\sigma = 47.2\pm 6.1$~pb
(when collinear improvement is included).
However, as it is well known~\cite{PI_SQ}, the inclusion of $\pi^2$ enhanced terms to all orders leads
to a much faster convergence, as shown in Fig.~\ref{fig:1}; this is especially manifest in the
collinear improved case, where we also observe that all the resummed results have overlapping uncertainty bands,
thereby suggesting a more reliable estimate of the scale uncertainty.

\section{Conclusions}

We have computed the three loop soft function for inclusive Higgs production in gluon fusion,
deriving it from the equivalence of the resummation as performed in dQCD and in SCET.
Our result agrees with the very recent computation of Ref.~\cite{Li:2014afw}.
We have used it to compute the cross section at NNLO+N$^3$LL\pr\ in SCET, improving on the result
of Ref.~\cite{Ahrens:2008nc}, pushing it to the same accuracy of the current state of the art dQCD result~\cite{Bonvini:2014joa}.
Also, we propose a minimal modification of the form of the resummed soft logarithms which allows
to include collinear contributions to the result.
We find that the inclusion of these contributions is important, and leads to a faster convergence
of the perturbative expansion, with more reliable uncertainty estimate from scale variation,
further improved if the resummation of $\pi^2$ enhanced terms is included.
The result is also compatible within the uncertainties with the NNLO+N$^3$LL\pr\ obtained in dQCD
in Ref.~\cite{Bonvini:2014joa}.
The numerical results are available through the public code \texttt{ResHiggs}, version 2.0 onwards~\cite{ResHiggs}.

\begin{acknowledgments}
We thank T.~Becher, S.~Forte, S.~Marzani and G.~Ridolfi for discussions,
and the anonymous referee for useful comments and suggestions.
This research is supported by an European Research Council Starting Grant ``PDF4BSM: Parton Distributions in the Higgs Boson Era''.
\end{acknowledgments}

\end{document}